\documentclass[aps,prb,twocolumn,showpacs,superscriptaddress,nolongbibliography]{revtex4-2}
\usepackage[colorlinks ,linkcolor=blue,anchorcolor=blue,citecolor=blue,urlcolor=blue]{hyperref}
\usepackage{graphicx}
\usepackage{epstopdf}
\usepackage{soul, color, xcolor}
\usepackage{bm}
\usepackage{siunitx} 
\usepackage{amsmath,amssymb,amsfonts}
\usepackage{appendix}

\def\be{\begin{equation}} \def\ee{\end{equation}}
\def\bea{\begin{eqnarray}} \def\eea{\end{eqnarray}}

\begin{document}

\title{Valley-contrasting Spin Textures in Janus Metal Phosphochalcogenides}

\author{Zeyu Yin}
\affiliation{Center for Quantum Transport and Thermal Energy Science,
Institute of Physics Frontiers and Interdisciplinary Sciences,
School of Physics and Technology, Nanjing Normal University, Nanjing 210023, China}

\author{Li Liang}
\affiliation{Key Laboratory of Quantum Materials and Devices of Ministry of Education, School of Physics, Southeast University,
Nanjing 211189, China} 

\author{Zhichao Zhou}
\affiliation{Center for Quantum Transport and Thermal Energy Science,
Institute of Physics Frontiers and Interdisciplinary Sciences,
School of Physics and Technology, Nanjing Normal University, Nanjing 210023, China}

\author{Xiao Li} 
\email{lixiao@njnu.edu.cn}
\affiliation{Center for Quantum Transport and Thermal Energy Science,
Institute of Physics Frontiers and Interdisciplinary Sciences,
School of Physics and Technology, Nanjing Normal University, Nanjing 210023, China}

\begin{abstract} 
Momentum-resolved spin textures and potential valley-contrasting physical properties in the momentum space are two intriguing characteristics of noncentrosymmetric materials, and they have broad applications in spintronics and valleytronics. The realization of diverse spin textures within a single material, along with their further coupling to the valley degree of freedom, is highly desirable. Via first-principles calculations, we investigate electronic properties of Janus MP$_2$S$_3$Se$_3$ monolayers, which exhibits distinct spin textures at different valleys. While Ising-type spin textures are located at $K_\pm$ valleys, the symmetry breaking from the Janus structure brings about a coexistence of Weyl-type and Rashba-type spin textures at $\Gamma$ valley. In addition to valley-contrasting spin textures, valley dependence also occurs in Berry-curvature-driven anomalous Hall currents and optical selectivity. Besides, energy differences between $\Gamma$ and $K_\pm$, as well as band gaps, are highly tunable by applied strain. These findings present an intriguing coupling between diverse spin textures and multiple valleys, and pave the way for designing advanced electronic devices that leverage spin and valley degrees of freedom.
\end{abstract} 
\maketitle

{\color{blue}\textit{Introduction.}}-- Spin texture in momentum space, with the spin direction closely tied to the wave vector, is generated by the spin-orbit coupling in noncentrosymmetric crystals, and has enormous applications in spintronics \cite{manchon2015new, Chen2021}. Compared with well-known Rashba-type and Dresselhaus-type spin textures, Ising-type and Weyl-type ones have been merely explored and have only a few material examples  \cite{Vasko1979, Bychkov1984, Nitta1997, Zhu2011, Xiao2012, Hirayama2015, Sakano2020, Gatti2020, liang2025coupling}.
In transition-metal chalcogenides, the Ising-type spin texture is found with valley-locked out-of-plane spin polarization, which renders the valley index a good quantum number insensitive to non-magnetic impurities, and also gives rise to Ising superconductivity against external magnetic field \cite{saito2016superconductivity,PhysRevLett.109.166602,delabarrera2018tuning, wang2019typeii}. 
The Weyl-type one exhibits radial spin-momentum locking \cite{weyl1929electron, krieger2024weyl}, which has been experimentally observed in chiral Tellurium \cite{PhysRevLett.125.216402}. 
The spin texture can induce nonequilibrium spin polarization along the applied current, distinct from Rashba-type one \cite{PhysRevB.89.195418,PhysRevLett.109.166602,feng2016spin,Sakano2020}. 
Recent studies on SnP$_2$Se$_6$ monolayer have revealed multiple spin textures at different valleys, i.e., Ising-type ones at $K_\pm$ valleys and Weyl-type one at $\Gamma$ valley \cite{liang2025coupling}. 
This is a remarkably scarce instance of both rare spin textures coexisting in a single material, which establishes SnP$_2$X$_6$ as an exceptional platform for exploring the coupling between spin textures and valley degree of freedom.

On the other hand, two-dimensional Janus materials have two asymmetric surfaces. The asymmetric structure breaks the horizontal mirror symmetry and generates a spontaneous internal electric field \cite{riis2018efficient, zhang2025giant, article}, resulting in physical phenomena that are absent in its symmetric parent structure \cite{PhysRevB.103.035414, PhysRevB.105.245420}. 
For example, in Janus MoGeSiP$_2$ monolayers, the broken mirror symmetry enables Rashba-type spin texture characterized by an orthogonal spin-momentum locking \cite{HUSSAIN2022169897, PhysRevB.97.235404, 929b-8kcy}. 
Janus materials can thus be utilized for exploring spin-orbit physics and designing spintronic devices. 
As to metal phosphochalcogenides, MP$_2$X$_6$ (M = Ge, Sn; X = S, Se), the introduction of a Janus structure is expected to endow this rising material family with richer spin textures and spintronic applications.

In this work, using first-principles calculations, we investigate atomic and electronic properties of Janus metal phosphochalcogenides, MP$_2$S$_3$Se$_3$ (M = Ge, Sn, Pb), with GeP$_2$S$_3$Se$_3$ as a typical example. 
The band structure calculations demonstrate the GeP$_2$S$_3$Se$_3$ monolayer has a semiconducting feature, with both the conduction and valence bands containing $K_\pm$ and $\Gamma$ valleys. 
Interestingly, the monolayer exhibits valley-contrasting spin textures. While $K_\pm$ valleys have Ising-type spin textures with opposite out-of-plane spins, the in-plane Weyl-type and Rashba-type ones simultaneously appear at $\Gamma$ valley, due to the Janus structure breaking both the mirror symmetry and rotational symmetry.
Valley-dependent Berry curvature and circular dichroism are also identified at $K_\pm$ valleys. 
Furthermore, the ratio between Weyl-type and Rashba-type spin textures at $\Gamma$ valley varies with different MP$_2$S$_3$Se$_3$ monolayers.
The band gaps and energy difference between $\Gamma$ and $K_\pm$ valleys are calculated to be tunable through the biaxial strain. 
The Janus metal phosphochalcogenides monolayer, with valley-contrasting spin textures and the coexistence of different spin textures at a single valley, offers a promising arena for studying physical phenomena associated with spin and valley degrees of freedom, and for designing high-performance spintronic and valleytronic devices.

\begin{figure}[htb]
\includegraphics[width=8.5 cm]{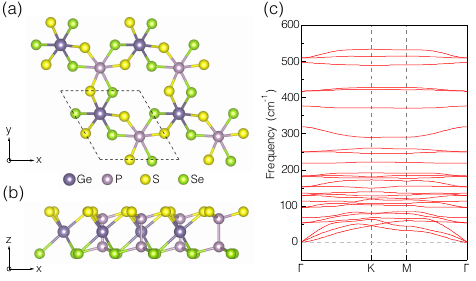}
\caption{Atomic structure and phonon spectrum of the GeP$_2$S$_3$Se$_3$ monolayer.
(a) Top view and (b) side view of the monolayer. 
Purple, pink, yellow, and green balls stand for Ge, P, S, and Se atoms, respectively.
(c) Phonon spectrum of the monolayer.
}

\label{fig01}
\end{figure}

{\color{blue}\textit{Crystal structures and phonon spectrum.}} --
Taking GeP$_2$S$_3$Se$_3$ monolayer as an example, we first study the atomic structure of two-dimensional metal phosphochalcogenides. 
Figs.\,\ref{fig01}(a) and (b) show the top and side views of the GeP$_2$S$_3$Se$_3$ monolayer, respectively. 
The monolayer has a hexagonal lattice, with one Ge$^{4+}$ ion and an anion cluster composed of two P, three S, and three Se atoms in a unit cell. 
In the anion cluster, the S and Se atoms are located above and below the vertical P-P bond, respectively, form a staggered conformation.
On the other hand, the Ge ion has an octahedral coordination with the six chalcogen atoms.

After full relaxation of the atomic structure via first-principles calculations, the in-plane lattice constant \(a\) of the GeP$_2$S$_3$Se$_3$ monolayer is computed to be 6.19 Å. 
The vertical distance between the uppermost and lowermost atomic layers gives a monolayer thickness of 3.34 Å. 
The optimized bond lengths are calculated to be 2.23 Å for the P–P bond, 2.05 Å for the P–S bond, and 2.22 Å for the P-Se bond, respectively.

By employing the finite displacement method, we calculate the phonon spectrum of the GeP$_2$S$_3$Se$_3$ monolayer. 
As depicted in Fig.\,\ref{fig01}(c), there are no notable imaginary frequencies within the whole Brillouin zone, indicating that the Janus structure is dynamically stable.

Furthermore, from the view of the crystal symmetry, the GeP$_2$S$_3$Se$_3$ monolayer has a point group of $C_3$. 
The horizontal mirror symmetry is absent in the monolayer, due to the Janus structure and the staggered conformation of the anion cluster. 
Compared with the recently studied SnP$_2$Se$_6$ monolayer \cite{liang2025coupling}, the Janus structure further breaks the twofold rotational symmetry with the horizontal rotational axis, which offers the potential for richer spin textures. 
On the other hand, the preserved threefold rotational symmetry around the z-axis plays an essential role in establishing possible valley-contrasting physical properties. 
The spin textures and valley-contrasting physics  will be elaborated subsequently.

\begin{figure}[htb]
\includegraphics[width=8.5 cm]{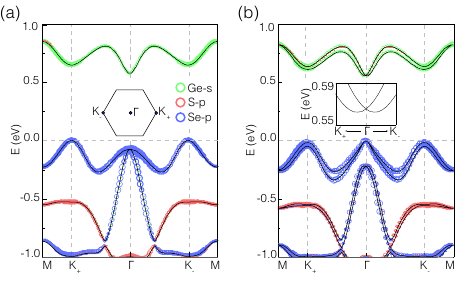}
\caption{Orbital-projected band structures of the GeP$_2$S$_3$Se$_3$ monolayer.
 (a) Band structure without the SOC. (b) Band structure with the SOC. 
 The orbital projections are demonstrated by a fat-band representation, with the radii of colored circles being proportional to orbital contributions. The valence band maxima are set to zero energy. The insets of (a) and (b) present the Brillouin zone of the monolayer with high-symmetry momentum points and magnified conduction bands in the vicinity of $\Gamma$ valley, respectively.
}
\label{fig02}
\end{figure}

{\color{blue}\textit{Band structrue with multiple valleys.}} --Fig.\,\ref{fig02} presents the band structures of the Janus GeP$_2$S$_3$Se$_3$ monolayer with orbital projections, using the Perdew-Burke-Ernzerhof functional \cite{Perdew1996}. 
To investigate the role of the spin-orbit coupling (SOC), the band structures are plotted without and with considering the SOC in Figs.\,\ref{fig02}(a) and (b), respectively. 
In Fig.\,\ref{fig02}(a), the non-relativistic band structure demonstrates that the monolayer is an indirect-band-gap semiconductor, with the conduction band minimum (CBM) at $\Gamma$ point and the valence band maxima (VBM) at both $K_+$ and $K_-$ points. 
The indirect band gap is computed to be 0.58 eV. 
Additionally, the band states at $K_+$ and $K_-$ points always remain degenerate besides the aforementioned VBM, which is protected by the time-reversal symmetry. 
Moreover, the conduction band has degenerate local minima at $K_\pm$, while the valence band has a local maximum at $\Gamma$. 
Therefore, there are both $K\pm$ and $\Gamma$ valleys for both conduction and valence bands. The sizes of the direct band gaps at $K_\pm$ valleys and $\Gamma$ valley are 0.65 eV and 0.66 eV, respectively. 
According to orbital projections in Fig.\,\ref{fig02}(a), it is seen that the conduction band is mainly contributed by the $s$ orbital of the Ge atom, whereas the valence band originates from $p$ orbitals of the Se atoms. 

When the SOC is considered in Fig.\,\ref{fig02}(b), the relativistic bands exhibit spin splitting and changes in the relative energy of different valleys.
It is seen that the valence band state at $\Gamma$ valley becomes higher than those at $K_\pm$ valleys, i.e., the VBM is shifted from $K_\pm$ to $\Gamma$. 
As to the conduction band, the CBM is still located at $\Gamma$ valley, but with a lateral momentum shift off the $\Gamma$ point, as illustrated in the inset of Fig.\,\ref{fig02}(b). 
The energy difference between the CBM and the conduction band states at $\Gamma$ point has a value of 3.23 meV, and the lateral momentum shift is 4.57$\times$10$^{-2}$ Å$^{-1}$. 
The value of the above energy difference is comparable to those in Rashba materials, e.g., BiAlO$_3$ and Janus SnSTe monolayer, indicating the possible appearance of a considerable Rashba-type or another SOC in the conduction band \cite{PhysRevB.93.245159, Bhat_2023}. 
In contrast, the momentum shift is not obvious at $\Gamma$ valley for the valence band. 
Besides, the SOC-induced spin splitting of the conduction band at $K_\pm$ valleys is computed to be 24 meV, while that of the valence band is 31 meV.
The direct band gap at $\Gamma$ and $K_\pm$ valleys are 0.56 eV and 0.76 eV, respectively. 
Moreover, the orbital projections on the conduction and valence bands are kept when the SOC is introduced, which confirms that the band splittings of both the valence and conduction bands arise from the spin involvement. 

Given that the Heyd-Scuseria-Ernzerhof (HSE) functional \cite{Heyd2003} can give more accurate band gaps than PBE functional, we then perform the HSE calculations of the GeP$_2$S$_3$Se$_3$ monolayer. 
The results are provided in the Supplementary Material (SM hereafter) \cite{SM}. As shown in SM, the band structure obtained by the HSE functional qualitatively resemble the PBE results, apart from larger band gaps. 
Without the SOC, the indirect band gap between the CBM at $\Gamma$ and VBM at $K_\pm$ is computed to be 1.29 eV. With the SOC, the direct band gap at $\Gamma$ valley is 1.33 eV.

\begin{figure}[htb]
\includegraphics[width=8.5 cm]{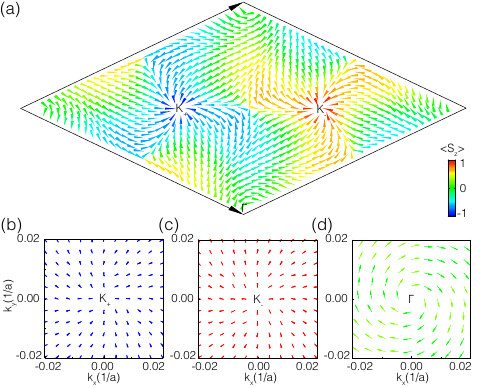}
\caption{Spin textures of the lower states of the spin-split conduction band of the GeP$_2$S$_3$Se$_3$ monolayer.
(a) The spin texture in a unit cell of the reciprocal lattice, with two vectors on the boundary denoting the unit vectors.
(b)-(d) The magenified spin textures at $K_+$, $K_-$ and $\Gamma$ valleys, respectively. The length and direction of arrows represent the magnitude and orientation of in-plane spin components, while out-of-plane spin components are coded by color.}
\label{fig03}
\end{figure}

{\color{blue}\textit{Valley-contrasting spin textures.}} -- To elucidate the types of spin splittings at different valleys, we investigate spin textures of the spin-split conduction band of the GeP$_2$S$_3$Se$_3$ monolayer.
Figs.\,\ref{fig03} (a) and (b)-(d) show the spin textures of the lower states of the spin-split conduction band in a unit cell of the reciprocal lattice and in the neighborhood of different valleys, respectively, while those of the upper states are presented in SM \cite{SM}. 

According to Figs.\,\ref{fig03} (a)-(c), it is found that the spin textures at $K_\pm$ valleys mainly consist of out-of-plane spin components. 
Specifically, the conduction band states are spin-down ones at $K_+$ valley and spin-up ones at $K_-$ valley, indicating valley-dependent Ising-type spin textures. The spin textures at $K_\pm$ valleys are similar to those in transition metal chalcogenides \cite{Xiao2012} and the SnP$_2$S$_6$ monolayer \cite{liang2025coupling}.   

In contrast, the spin textures at $\Gamma$ valley are dominantly contributed by in-plane components, as shown in Figs.\,\ref{fig03} (a) and (d).
The in-plane spins are found to be neither perpendicular nor parallel to the wave vector. In other words, the spins include both perpendicular and parallel components.
Since Weyl-type and Rashba-type spin textures feature parallel and perpendicular spin-momentum locking, respectively, the spin textures at $\Gamma$ valley are a coexistence of both Weyl-type and Rashba-type ones. 
This is different from that of the SnP$_2$Se$_6$ monolayer, where only the Weyl-type spin texture exists at $\Gamma$ valley \cite{liang2025coupling}. 
Furthermore, the ratio of Weyl-type to Rashba-type spin texture at $\Gamma$ valley is calculated to be approximately 0.20 by obtaining the ratio of spin components parallel and perpendicular to the wave vector. 
Besides, compared with the lower ones, the upper states of the spin-split conduction band are found to have opposite spins at both $\Gamma$ and $K_\pm$ valleys (see SM) \cite{SM}.  
According to the above analysis, the GeP$_2$S$_3$Se$_3$ monolayer has not only valley-contrasting spin textures, but also multiple spin textures at a single valley.  

The aforementioned spin textures can be well described by the Hamiltonians with different types of SOC.  
The Ising-type spin texture at $K_\pm$ valleys corresponds to Ising-type SOC, which is given as $\tau\sigma_z$. Here, the valley index $\tau$ = $\pm$1 denote $K_\pm$ valleys, respectively, and $\sigma_z$ is the out-of-plane spin Pauli matrix. 
When $\tau$ = $\pm$1, the lower states of the Hamiltonian have down and up spins, respectively, which match the cases of $K_\pm$ valleys in Figs.\,\ref{fig03}(b) and (c). 
On the other hand, as to the coexistence of Weyl-type and Rashba-type spin textures at $\Gamma$ valley, the SOC Hamiltonian can be written as $\alpha(k_x\sigma_x+k_y\sigma_y)+\beta(k_x\sigma_y-k_y\sigma_x)$. Here, the first and second terms correspond to Weyl-type and Rashba-type SOC, respectively, with $\alpha$ and $\beta$ as corresponding strengths. $k_{x,y}$ are in-plane wave vectors, and $\sigma_{x,y}$ are in-plane spin Pauli matrices. 
The Weyl-type and Rashba-type SOC in the Hamiltonian provide in-plane spin components parallel and perpendicular to the wave vector, respectively. The coexistence of two types of SOC enables the spins neither parallel nor perpendicular to the wave vectors, as illustrated in Fig.\,\ref{fig03}(d). Additionally, since the ratio of $\alpha$ to $\beta$ is equal to that of the parallel spin component to the perpendicular one (see SM) \cite{SM}, the ratio is thus 0.20 for the GeP$_2$S$_3$Se$_3$ monolayer.

The spin textures can be understood from the perspective of symmetry. 
As mentioned above, the Janus structure of the GeP$_2$S$_3$Se$_3$ monolayer breaks both the horizontal mirror and twofold rotational symmetries, reducing the point group of the monolayer from $D_3$ to $C_3$. The little groups at both $\Gamma$ and $K_\pm$ are $C_3$ as well. The $C_3$ symmetry at $K_+$ excludes in-plane spin components. Further considering the time-reversal symmetry that relates $K_+$ and $K_-$, the out-of-plane spins at $K_\pm$ are opposite, leading to valley-contrasting Ising-type spin texture at $K_\pm$. 
On the other hand, the $C_3$ symmetry also allows for the appearances of the Weyl-type, Rashba-type spin textures or their coexistence in the vicinity of corresponding momentum points, which occur at $\Gamma$ valley of the conduction band.

Additionally, the spin textures of the valence band are also calculated and provided in the SM \cite{SM}. Compared with the conduction band, in-plane spin components decrease at $\Gamma$ valley for the valence band, while the out-of-plane components increase, which may result from different orbital contributions to the valence and conduction bands (see Figs.\,\ref{fig02}). As to $K_\pm$ valleys of the valence band, the spin textures are still valley-contrasting Ising-type ones.

\begin{figure}[htb]
\includegraphics[width=8.5 cm]{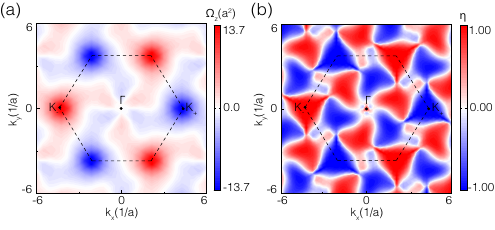}
\caption{ Valley-contrasting transport and optical properties of the GeP$_2$S$_3$Se$_3$ monolayer. 
(a) The distribution of the Berry curvature in the momentum space.
(b) The degree of circular polarization in the momentum space. The Brillouin zone is bounded by dashed lines.
}
\label{fig04}
\end{figure}

{\color{blue}\textit{Berry curvature and circular dichroism.}} -- We further study the Berry curvature and circular dichroism of the GeP$_2$S$_3$Se$_3$ monolayer. 
Fig.\,\ref{fig04}(a) shows the momentum-resolved Berry curvature, $\Omega_z(\textbf{\textit{k}})$. 
It is found that significant and opposite Berry curvature appear at $K_\pm$ valleys and the Berry curvature vanishes at $\Gamma$ valley.
As a result, when the monolayer is under an in-plane electric field and carrier doping, opposite Berry-curvature-driven anomalous Hall currents with opposite spins arise from the $K_\pm$ valleys, resulting in both valley and spin Hall effects.

Fig.\,\ref{fig04}(b) displays the momentum-resolved degree of circular polarization, $\eta(\textbf{\textit{k}})$, of direct optical transition between the upper valence band state and lower conduction band state. 
It is found that $\eta(\textbf{\textit{k}})$ acquires finite values with opposite signs in the vicinity of $K_\pm$ valleys, and it reaches $\pm$1 exactly at $K_\pm$ points. 
This means that left- and right-circularly polarized lights are exclusively absorbed at $K_+$ and $K_-$, respectively. 
Consequently, anomalous Hall currents from $K_\pm$ valleys can also be selectively generated by leveraging this circular dichroism \cite{cao2012}. 
Therefore, besides the valley-contrasting spin textures mentioned above, the $K_\pm$ valleys in the GeP$_2$S$_3$Se$_3$ monolayer possess both opposite Berry curvatures and optical selectivities, allowing valley-dependent optoelectronic responses.

{\color{blue}\textit{Modulation of valley-related properties.}} -- Besides the Janus GeP$_2$S$_3$Se$_3$ monolayer, first-principles calculations are also performed to investigate atomic and electronic properties of other MP$_2$S$_3$Se$_3$ (M = Sn, Pb) monolayers with similar crystal structures, which are shown in Table \ref{table1}. In the calculations, the SnP$_2$S$_3$Se$_3$ monolayer has an indirect band gap with the CBM and VBM being located at $\Gamma$ and $K_\pm$ valleys, respectively, while the PbP$_2$S$_3$Se$_3$ monolayer has a global direct band gap at $K_\pm$ valleys, which are different from the GeP$_2$S$_3$Se$_3$ monolayer. More interestingly, SnP$_2$S$_3$Se$_3$ and PbP$_2$S$_3$Se$_3$ monolayers exhibit different ratios of the Weyl-type to Rashba-type spin textures at $\Gamma$ valley from the GeP$_2$S$_3$Se$_3$ monolayer, with values of 0.34 and 0.54, respectively. This indicates that the ratio of these two spin textures can be tuned by the replacement of the metal atom. The tunable ratio is likely to result in different consequences in spin-related phenomena, e.g., current-induced spin polarization. Given that current-induced spin polarizations are parallel and perpendicular to the applied current for Weyl-type and Rashba-type spin textures, respectively, the orientation of non-equilibrium spins tends to gradually align with the direction of applied current as the ratio increases, which is worth further study. 

 \begin{table}[h]
    \centering
    \caption{Atomic and electronic properties of MP$_2$S$_3$Se$_3$ monolayers. The properties include the lattice constant ($a$), the band gap with the SOC, and the ratio of Weyl-type to Rashba-type spin textures (W/R). The band gaps outside and inside the parentheses are calculated using PBE and HSE functionals, respectively.}
    \begin{ruledtabular}
    \begin{tabular}{cccccc}
        & $a$ (Å) &  Band gap (eV) & W/R \\
    \hline \\[-2.0 ex]
        GeP$_2$S$_3$Se$_3$ & 6.19 & 0.563 (1.334) & 0.20 \\
        SnP$_2$S$_3$Se$_3$ & 6.32 & 0.814 (1.687)  & 0.33\\
        PbP$_2$S$_3$Se$_3$ & 6.41 & 0.006 (0.649)  & 0.54\\
    \end{tabular}
    \end{ruledtabular}
\label{table1}    
\end{table}

\begin{figure}[htb]
\includegraphics[width=8.5 cm]{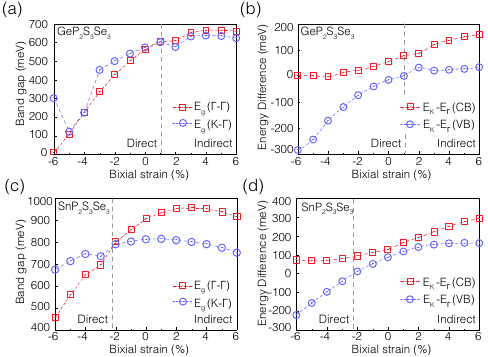}
\caption{Evolutions of band gaps and relative energies between different valleys with applied strain.
(a) The evolutions of the direct band gap at $\Gamma$ valley and the indirect band gap between the valence band state at $K_\pm$ and the conduction state at $\Gamma$ of the GeP$_2$S$_3$Se$_3$ monolayer.
(b) The evolutions of the relative energies between $K_\pm$ and $\Gamma$ valleys for both conduction and valence bands. 
(c) and (d) are corresponding evolutions for the SnP$_2$S$_3$Se$_3$ monolayer.}
\label{fig05}
\end{figure}

On the other hand, we investigate the influence of biaxial strain on band gaps and relative energies between $K_\pm$ and $\Gamma$ valleys in GeP$_2$S$_3$Se$_3$ and SnP$_2$S$_3$Se$_3$ monolayers, as depicted in Figs.\,\ref{fig05}(a)-(d).
Fig.\,\ref{fig05}(a) shows the strain-dependent evolutions of the direct band gap at $\Gamma$ valley and the indirect band gap between the valence band state at $K_\pm$ and the conduction state at $\Gamma$ of the GeP$_2$S$_3$Se$_3$ monolayer. It is found that the direct band gap gradually increase under the strain from $-$6\% to +6\%, while the indirect band gap has a non-monotonic variation. 
Moreover, there is a transition from a global direct band gap to an indirect one at about +1\% strain.
The transition is due to the change of relative energies between $K_\pm$ and $\Gamma$ valleys, as illustrated in Fig.\,\ref{fig05}(b).
For the conduction band, the energy difference between $K_\pm$ and $\Gamma$ is smaller than 4 meV for the strain from $-$6\% to $-$4\%, and then the energy difference enhances as the strain varies from $-$3\% to +6\%.
As to the valence band, the VBM shifts from $\Gamma$ to $K_\pm$ as the strain increases, with about +1\% strain as a critical point, which leads to the transition in Fig.\,\ref{fig05}(a). 
Besides, the effect of strain on the band gaps and the relative energies between different valleys of the SnP$_2$S$_3$Se$_3$ monolayer exhibit similar trends with the critical point at about $-$2\% strain. Therefore, these characteristics of the band structures are highly tunable via applied strain.  


{\color{blue}\textit{Summary}} -- To conclude, taking the GeP$_2$S$_3$Se$_3$ monolayer as a representative example, we employ first-principles calculations to investigate atomic and electronic properties of the MP$_2$S$_3$Se$_3$ family. 
The GeP$_2$S$_3$Se$_3$ monolayer is found to have valley-contrasting spin textures, with opposite Ising-type ones at $K_\pm$ valleys, and the coexistence of both Weyl-type and Rashba-type ones at $\Gamma$ valley. 
Multiple spin textures in the GeP$_2$S$_3$Se$_3$ monolayer, particularly at $\Gamma$ valley, indicate that the monolayer may possess distinct consequences and high tunabilities in spin-related phenomena, compared with materials that have only one single type of spin texture. 
Furthermore, both the Berry curvature and the chiral optical selectivity also exhibit pronounced valley-contrasting behaviors in the monolayer. 
Additionally, the above compelling characteristics can be generalized to another MP$_2$S$_3$Se$_3$ monolayer. The biaxial strain is also utilized as an effective external stimulus for modulating band gaps and relative energies between different valleys. 
Our study reveals that Janus metal phochalcogenides monolayers can serve as a versatile platform for exploring spin‑ and valley‑related physical properties, and paves pathways for the development of advanced spintronic and valleytronic devices.

{\color{blue}\textit{Acknowledgments}} -- 
We are supported by the National Natural Science Foundation of China (Nos.\,12374044, 11904173, 12004186). 

\appendix

%

\providecommand{\noopsort}[1]{}

\end{document}